\documentclass[aps,prb,twocolumn,superscriptaddress]{revtex4-1}
\usepackage{graphicx}
\usepackage{dcolumn}
\usepackage{bm}
\usepackage{amsmath}
\usepackage{amssymb}
\usepackage[usenames,dvipsnames]{xcolor}

\begin{document}


\title{Berry curvature dipole current in transition metal dichalcogenides family}

\author{Jhih-Shih You}
\email{jhihshihyou@gmail.com }
\affiliation{Department of Physics, Harvard University, Cambridge, Massachusetts 02138, USA}
\affiliation{Institute for Theoretical Solid State Physics, IFW Dresden, Helmholtzstr. 20, 01069 Dresden, Germany}
\author{Shiang Fang}
\email{shiangfang913@gmail.com }
\affiliation{Department of Physics, Harvard University, Cambridge, Massachusetts 02138, USA}

\author{Su-Yang Xu}
\affiliation{Department of Physics, Massachusetts Institute of Technology,
Cambridge, Massachusetts 02139, USA}

\author{Efthimios Kaxiras}
\affiliation{Department of Physics, Harvard University, Cambridge, Massachusetts 02138, USA}
\affiliation{John A. Paulson School of Engineering and Applied Sciences, Harvard University, Cambridge, Massachusetts 02138, USA}
\author{Tony Low}
\email{tlow@umn.edu}
\affiliation{Department of Electrical and Computer Engineering, University of Minnesota, Minneapolis, Minnesota 55455, USA}

\date{\today}

\pacs{72.15.-v, 73.22.?f, 73.43.-f,03.65.Vf}


\begin{abstract}
We study the quantum nonlinear Hall effect in two-dimensional materials with time-reversal symmetry.
When only one mirror line exists, a transverse charge current occurs in second-order response to an external electric field, as a result of the Berry curvature dipole in momentum space. Candidate 2D materials to observe this effect are two-dimensional transition-metal dichalcogenides~(TMDCs).
First we use an   {\it  ab initio} based tight-binding approach to demonstrate that monolayer $T_d$-structure TMDCs exhibit a finite Berry curvature dipole.
In the $1H$ and $1T'$ phase of TMDCs, we show the emergence of finite Berry curvature dipole with the application of strain and  electrical displacement field respectively. 
 \end{abstract} 
\maketitle


 TMDCs~\cite{Radisavljevic2011,Wang2012,Geim2013,Butler2013} have
lately attracted considerable attention  because of their rich physics, such as   charge density
wave~\cite{Wilson2001,Ritschel2015,Tsen2015}, superconducting phase~\cite{Ye2012}, two dimensional
(2D) quantum spin Hall~(QSH) state~\cite{Qian2014,Fei2017,Tang2017,Jia2017,Wu2018}  and Weyl semi-metal states~\cite{Soluyanov2015},
among other phenomena. Recently numerous studies have demonstrated new physical properties
in a monolayer~(ML) of TMDCs that may be different from those in bulk.  For example, molybdenum disulfide~($\rm{MoS_2}$) exhibits an induced indirect to direct band-gap transition
from $2H$-structure to
its ML $1H$-structure, together with an enhancement of
the luminescence quantum yield in comparison with the
$\rm{MoS_2}$ bulk~\cite{Splendiani2010,Mak2010}. The $1H$-structure
 which does not possess  a center of inversion, in contrast to the $2H$-structure, allows
optical control of valley degrees of freedom~\cite{Zeng2012}.

 \begin{figure}[t]
\centering
\includegraphics[width=8.0cm]{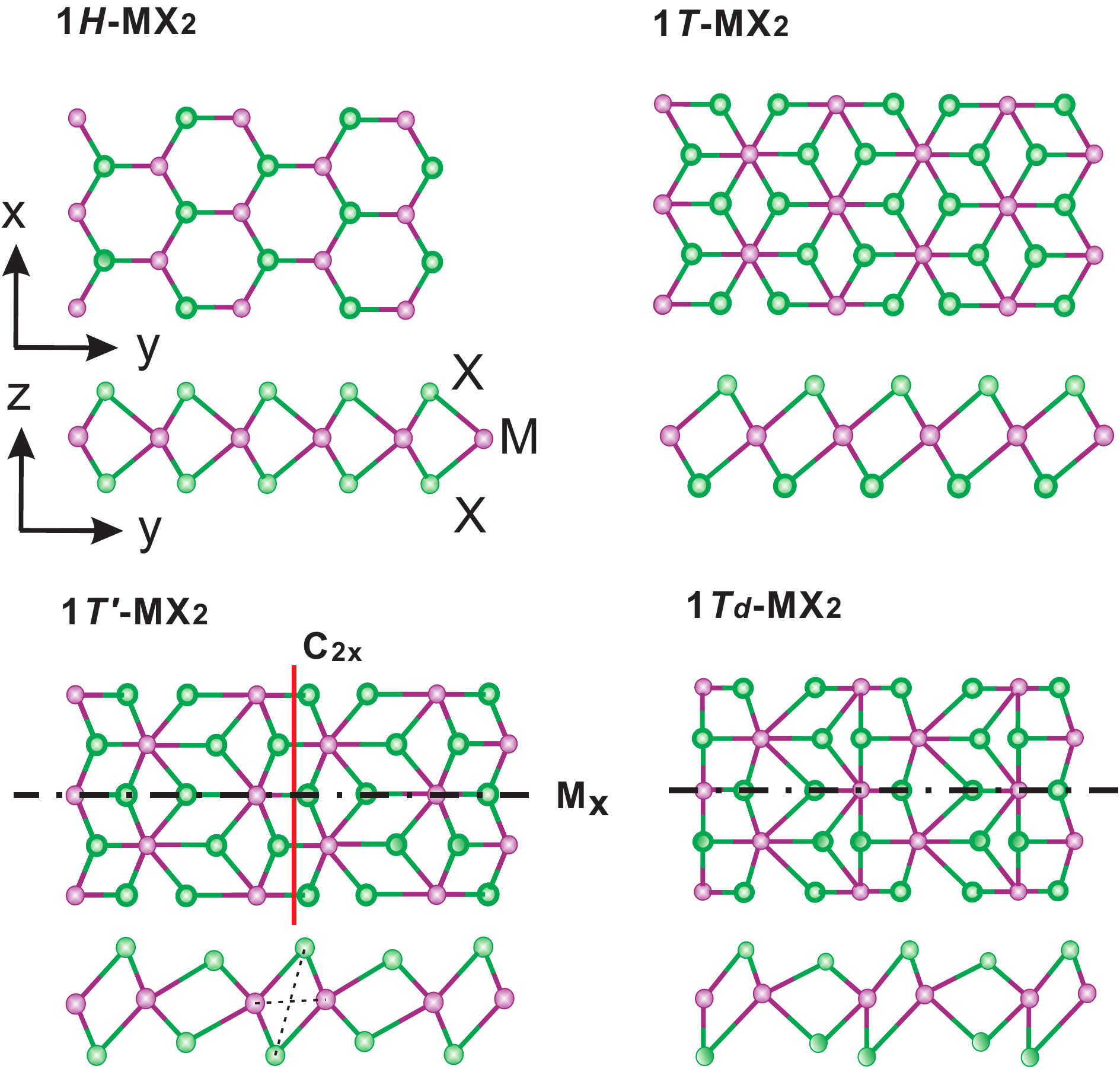}
\caption{Crystal structures of $1H,$ $1T,$ $1T'$ and $1T_d$ monolayer transition metal dichalcogenides $\rm{MX_2}$~(M = Mo, W and X = S, Se, Te). The red solid line represents a screw rotate symmetry $C_{2x}$ axis which involves a $180^{\circ}$ rotation about $\hat{x}$ and a half lattice-constant translation along $\hat{x}.$ The dashed-dotted line represents the mirror symmetry $M_x.$ The combination of  $C_{2x}$  and $M_x$  symmetries leads to the inversion symmetry for the $1T'$ structure. The $1T_d$ phase has only $M_x$ symmetry.
}\label{fig_lattice}
\end{figure}

Generally, the
electronic properties of monolayer TMDCs  with chemical
composition $\rm{MX_2}$~( M= Mo,
W; and X=  S, Se, Te) are directly related to their crystal structure, which includes the $1H$~( $P\bar{6}m2$ ), $1T$ ( $P\bar{3}m2$), $1T'$~( $P2_1/m$), and $T_d$~( $P1m1$) crystal structures, shown in Fig.~\ref{fig_lattice}.
$1H-\rm{MX_2} $ is a semiconductor with a direct band gap in the range of visible
light~(1 to  2 eV)~\cite{Qin1991,Radisavljevic2011,Fang2015} and the $1T$ structure is metallic~\cite{Kappera2014}. Topological phases occur in $1T'$
structures since the global
properties of electronic wavefunctions exhibit a non-trivial topology, theoretically predicted~\cite{Qian2014} and experimentally confirmed~\cite{Fei2017,Tang2017,Jia2017,Wu2018} to be a 2D quantum spin Hall state.

The local curvature of the wavefunction, defined
as the Berry curvature~(BC)~\cite{Nagaosa2010,Xiao2010}, is a geometrical property of the
Bloch energy band. The finite Berry curvature  reveals linear response, such as anomalous Hall conductivity~\cite{Nagaosa2010}, and nonlinear response, such as  circular photogalvanic effect~(CPGE) and
nonlinear Hall effect~\cite{Deyo2009,Moore2010,Low2015,Sodemann2015,Eginligil2015,deJuan2017,Tsirkin2018,Zhang2018,Quereda2018,Xu2018,Facio2018,Shi2018}.
In TMDCs, the BC at the two valleys  takes opposite values, giving rise to
bulk topological charge neutral valley current~\cite{Xiao2012,Mak2014,Lensky2015}, as a linear response.
For nonlinear response, the semiclassical approach has been used
to describe the intra-band contributions to transverse current at both zero frequency and
second harmonic generation in terms of the dipole moment of the BC in momentum space~\cite{Sodemann2015}. In the
dc limit, the photocurrent can remain finite as a transverse
Hall-like current~\cite{Sodemann2015,Xu2018}.

In this work, we study the emergence of the non-linear Hall current induced by the
 BC dipole in various  2D TMDC structures, using \textit{ab initio} calculations combined with a semiclassical
approach. Monolayer TMDCs with low crystalline symmetry are expected to be excellent candidates for observing these quantum non-linear effects~\cite{Sodemann2015}.
Our study quantitatively reveals the finite BC dipole in the
$T_d$-structure monolayer TMDCs, where only one mirror line survives. For the $1H$ structure, uniaxial strain can break the underlying $C_{3v}$ symmetry which otherwise ensures vanishing nonlinear response. In addition, we evaluate the Berry dipole of strained TMDC.
Lastly, we show that application of an out-of-plane  electrical displacement field induces a strong BC dipole which is absent in $1T'$ monolayer TMDCs.
Our results represent the first numerical demonstration of non-linear current in 2D TMDCs, which can
be controlled by mechanical and/or electrical means.
The proposed strategies of a tunable BC dipole also apply to a wide range of other
 two dimensional materials, such as
hexagonal boron nitride~\cite{Golberg2010} and black phosphorus~\cite{Morita1986,Li2014,Liu2014}.

In the
presence of an external electric field, the carrier velocity
contains contributions from the
group velocity of the electron wave and from the anomalous transverse term due to the BC, given by
$-\frac{e}{\hbar} \vec{E} \times \Omega_{z}^{(n)}(\vec{k}). $ Here
the BC for the electronic Bloch states of the
nth band is defined as~\cite{Xiao2010,Fuchs2010,Xu2014}
\begin{small}
\begin{eqnarray}
\Omega_{z}^{(n)}(\vec{k})&=& \mathrm{i} \hat{z}\cdot (\nabla_{\vec{k}} {u^{(n)}_{\vec{k}}} ^*)\times(\nabla_{\vec{k}}u^{(n)}_{\vec{k}}
)\\
&=& -2 \big(\frac{\hbar}{e}\big)^2 \sum_{n\neq n'}\frac{\textmd{Im}\langle u^{(n)}_{\vec{k}} |P_x(\vec{k})|u^{(n')}_{\vec{k}}\rangle\langle u^{(n')}_{\vec{k}}|P_y(\vec{k})|u^{(n)}_{\vec{k}}\rangle}{[\epsilon^{(n)}_{\vec{k}}-\epsilon^{(n')}_{\vec{k}}]^2}\nonumber\\
\end{eqnarray}
\end{small}
where $\epsilon^{(n)}_{\vec{k}}$ and $|u^{(n)}_{\vec{k}} \rangle$ are eigenvalues and eigenfunctions of the Hamiltonian $\hat{H}_{\vec{k}}$, respectively,
at the momentum $\vec{k}$ and  $P_i(\vec{k})=(e/\hbar)\partial \hat{H}_{\vec{k}}/\partial k_i$ is the current operator. The BC is
analogous to an effective magnetic field in the momentum space. Within linear response theory,
the integral of the BC over the entire Brillouin
zone gives rise to a transverse conductivity, which is simply
given by
$\sigma_{xy,n}= e^2/\hbar\int d^2 \vec{k} f^{(n)}_0(\vec{k}) \Omega_{z}^{(n)}(\vec{k}) ,$  where $  f^{(n)}_0(\vec{k})$ is the equilibrium Fermi-Dirac distribution function for $n$th band. The transverse conductivity is zero
for a time-reversal-invariant system, since states at $\vec{k}$ and $-\vec{k}$
are equally occupied and time-reversal symmetry requires that $\Omega_{z}^{(n)}(\vec{k})=-\Omega_{z}^{(n)}(-\vec{k}).$
When the system is driven out-of-equilibrium, a net transverse current can survive as the second order response to the
electric field.
The combination of time reversal~(TR) and inversion symmetry restricts $\Omega_{z}^{(n)}(\vec{k})=0$ over the entire Brillouin
zone. Thus, for a
TR invariant system, inversion symmetry breaking is necessary to generate a finite BC.

We first examine the nature of the non-linear current with a symmetry analysis. In the
presence of a driving in-plane electric field,
$E_k(t)=\Re \{\mathcal{E}_k \mathrm{e}^{\mathrm{i}\omega t}\},$
the non-linear current is written as $j_i=j^{(0)}_i+j^{(2\omega)}_i \mathrm{e}^{2\mathrm{i}\omega t},$ where the dc and second harmonic generated currents are described by the second-order susceptibility tensor as
$j^{(0)}_i = \chi_{ijk} \mathcal{E}^*_j  \mathcal{E}_k$ and $j^{(2\omega)}_i = \chi_{ijk} \mathcal{E}_j  \mathcal{E}_k,$ respectively.
The tensor
indices $i, j, k$ span the 2D sample coordinates $x,y$.  $\chi_{ijk} $ respects the symmetry of the crystal lattice.
The presence of $M_x$ mirror symmetry forces $\chi_{ijk} $
 to be zero if any $\chi_{ijk} $ contains an odd number of the index $x.$  All tensor components identically vanish in the presence
of inversion symmetry. Thus, for a crystal breaking inversion but preserving the mirror symmetry $M_x,$ $\chi_{yxx}\neq 0,$  demonstrating that a Hall-like transverse current can occur in second-order
response to an external electric field $\mathcal{E}_x.$

The expression of nonlinear currents has been theoretically obtained
within the semiclassical Boltzmann transport theory for a single band~\cite{Low2015,Sodemann2015}.  Up to second order in the  driving electric
field, previous theoretical works showed that in the case  involving only the intra-band process, a nonlinear Hall-like current density is expressed as
\begin{eqnarray}\label{current}
\vec{j}^{(0)}&=&\frac{e^3 \tau}{2 \hbar^2 (1+\mathrm{i}\omega \tau)} \hat{z}\times \vec{\mathcal{E}}^* (D \cdot\vec{ \mathcal{E} } )\\\label{current1}
\vec{j}^{(2\omega)}&=&\frac{e^3 \tau}{2 \hbar^2 (1+\mathrm{i} \omega \tau)} \hat{z}\times \vec{\mathcal{E}} (D \cdot\vec{ \mathcal{E} } )
\end{eqnarray}
with
\begin{eqnarray}\label{dipole}
D_i &=&   \sum_n\int d^2 \vec{k} f^{(n)}_0 (\vec{k}) [\partial_{k_i} \Omega_{z}^{(n)}(\vec{k})]\nonumber\\
&=& - \sum_n \int d^2 \vec{k} [ \partial_{k_i} f^{(n)}_0 (\vec{k})] \Omega_{z}^{(n)} (\vec{k})
\end{eqnarray}
where $\tau$
is the relaxation time. The  nonlinear dc current is proportional to the  dipole moment of the
BC over the occupied states.
In fact, according to Eq.~\eqref{dipole} the
non-linearity of these currents is associated with a
``Fermi-surface'' contribution, that is, only
states near the Fermi surface can contribute to the
integral in the low temperature limit.
 The
largest symmetry of a 2D crystal that allows for a non-vanishing BC dipole is a single mirror line~\cite{Sodemann2015}.
Combining Eq.~\eqref{dipole} with the fact that
$\Omega_{z}^{(n)}(k_x, k_y) = -\Omega_{z}^{(n)}(-k_x, k_y)$ enforced by the mirror plane
$M_x$, it is evident that $D_x\neq 0$ and $D_y= 0.$
Consequently, according to
Eq.s~\eqref{current} and ~\eqref{current1}, when the driving electric field $E_k(t)=\mathcal{E}_k $ is aligned with
the direction of the Berry curvature dipole vector $D_x$,  we obtain the dc current density $j= 2 {j}^{(0)}=3698.1 (A/m^2)\times  D_x\sim10^{-7} (A/m),$ where we choose $\tau=10^{-12} s,$ $D_x=1 \mathrm{\AA}$ and $ \mathcal{E}_x=100 V/m.$ 

\begin{figure}[t]
\centering
\includegraphics[width=8.5cm]{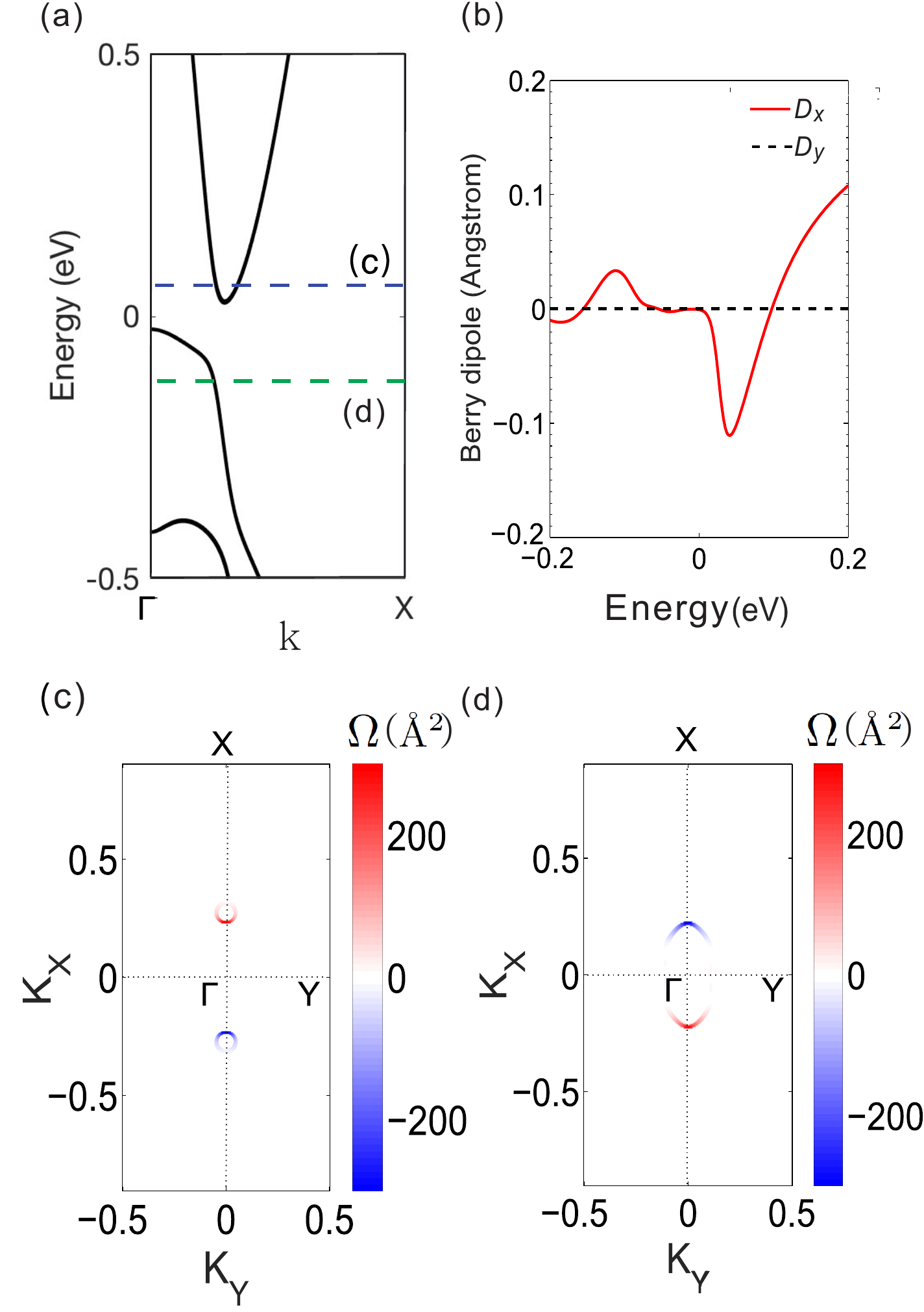}
\caption{(a) Band structure along  $\Gamma-$X  for the monolayer $T_d$-structure $\rm{WTe_2}$. The global band gap is around $0.05$ eV. The weak inversion breaking induces a tiny spin splitting near the bottom of the conduction band. The blue and green dashed lines correspond to $0.05$ eV and $-0.1$ eV at which the Berry curvature is shown in (c) and (d), respectively.   (b)  Berry curvature dipole $D_x$ and $D_y.$ (c) and (d) Berry curvature at $0.05$ eV and $-0.1$ eV, respectively.  
}\label{fig_1}
\end{figure}

\begin{figure}[t]
\centering
\includegraphics[width=9.0cm]{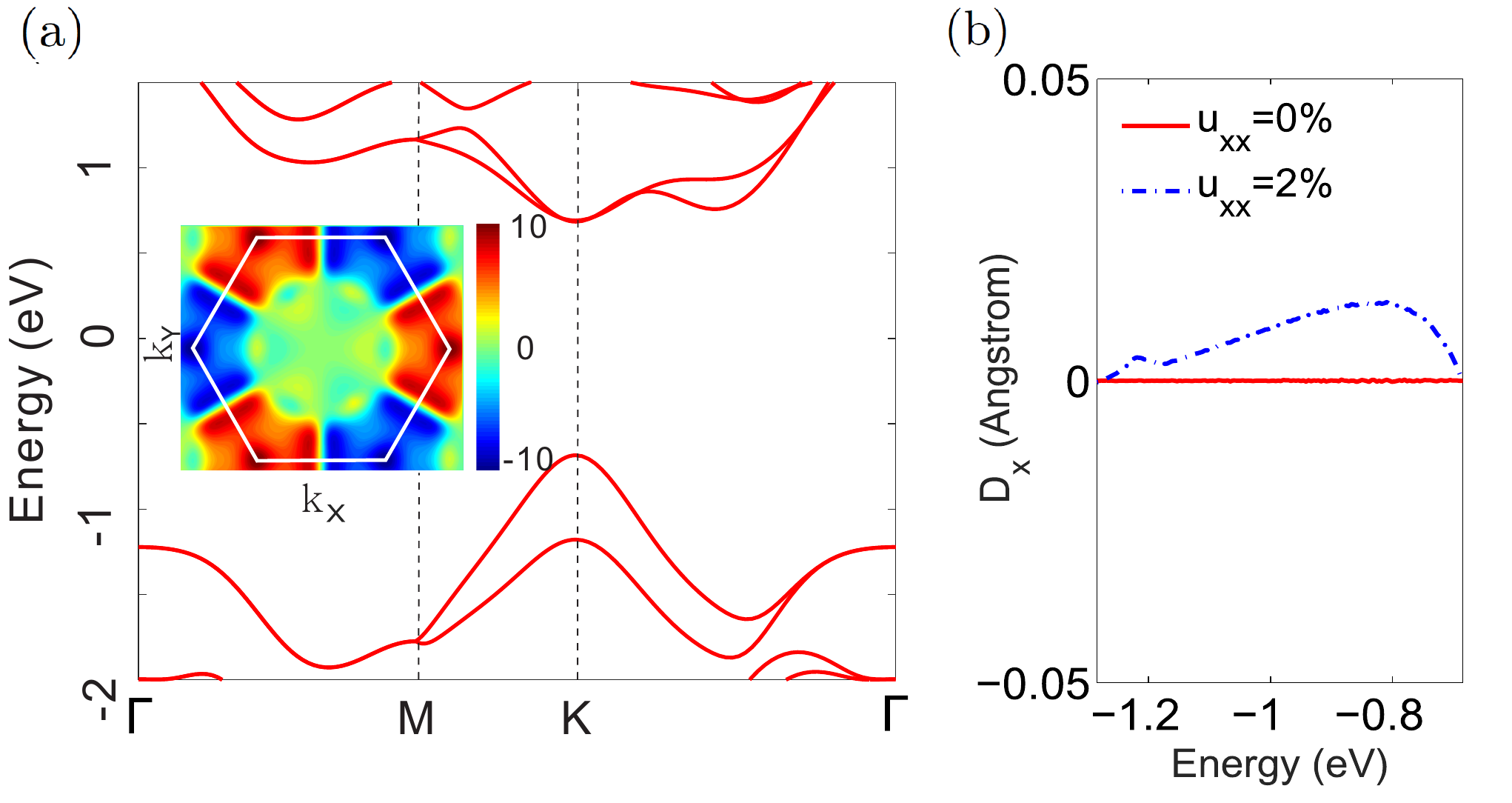}
\caption{(a) Band structure along  $\Gamma-$M$-$K$-\Gamma$  for the unstrained $H$-structure $\rm{WSe_2};$  the inset shows the Berry curvature contributed from the top two valance bands. The Brillouin zone is presented in white solid line. (b) Berry curvature dipole  $D_x$ of the monolayer $H$-structure $\rm{WSe_2}$ with strain $u_{xx}=0$~(red line) and $u_{xx}=2\%$~(blue dotted line) when $u_{yy}=u_{xy}=0$. $D_y=0$ for all strains.
}\label{fig_2}
\end{figure}

The BC dipole is tied to the underlying crystal
structure of monolayer TMDC.
 The most studied polymorphic
structures of pristine monolayer TMDCs are $1H$, $1T$, and $1T'$~\cite{Heising1999,Eda2012}, shown in Fig.~\ref{fig_lattice}.  The
 energetically favorable $1H-\rm{MX_2} $ layer is built from two hexagonal
lattices of X atoms and  an intercalated hexagonal plane of M  atoms  forming a simple ABA Bernal stacking with  $P\bar{6}m2$ space-group symmetry. In a monolayer
unit of $H$ structure, the inversion
symmetry is explicitly broken.
The lack of an inversion center in TMDCs produces substantial local
BCs near the $K,K'$ valleys. Due to time-reversal symmetry, the curvature at the two valleys has opposite sign,
which
implies counter-propagating currents
that persist even when the system is in equilibrium. Due to exact cancellation from
the two valleys, these
transverse currents are charge neutral to linear
response in an applied external electric field. The nonlinear contribution could still exist when the samples  have just one mirror symmetry.   2D materials like monolayer TMDCs in the $H$-structure  do not have currents in nonlinear response to the electric field because their $C_{3v}$ symmetry forces the BC dipole to vanish.

The $1T-\rm{MX_2} $ layer forms a rhombohedral ABC stacking phase with $P\bar{3}m2$ space group. In this structure, the transition
metal atoms are octahedrally coordinated.   DFT calculations show  that the
free-standing $1T$ structure is typically unstable and undergoes Peierls distortion in one direction to form a $2 \times 1$ reconstruction, where the distorted M atoms
form 1D zigzag chains~\cite{Kan2014},
referred to as the $T'$ structure.
Inversion symmetry is present in both the $T$- and $T'$-structure. The inversion-symmetric $1T'$ structure consists of two independent symmetries, the mirror symmetry $M_x$
and the two-fold screw rotational symmetry $C_{2x}.$ 
Due to the inversion symmetry and time-reversal symmetry,  $\Omega_{z}^{(n)}(\vec{k})=0$ over the entire Brillouin
zone for  the $1T'$-structure.

A  candidate material to observe the quantum non-linear Hall effect is monolayer $\rm{WTe_2},$ whose structure $T_d$ deviates slightly from the widely-studied $1T'$ structure.
  In  the $T_d$ phase, the $M_x$ mirror symmetry  is preserved but the
$C_{2x}$ symmetry is weakly broken. As a result, the $T_d$ structure  with symmetry space group $P1m1$  actually breaks inversion symmetry and allows a non-zero BC dipole to exist.
We perform  \textit{ab initio} density-functional theory~(DFT) calculations using the Vienna Ab
initio Simulation Package~(VASP)~\cite{vasp1,vasp2} with a minimal basis based on a transformation of the Kohn-Sham density functional theory Hamiltonian to a basis of maximally
localized Wannier functions~\cite{Mostofi2008} as
implemented in the Wannier90 code.  A slab geometry is
employed to model single or double layers with a $20 \mathrm{\AA}$
vacuum region between periodic images to minimize the
interaction between slabs. 
For TMDC materials, the relevant
states consist of seven valence bands and four conduction
bands, which are hybrids of metal $d$ orbitals and chalcogen
$p$ orbitals. Therefore, the Wannier
projections to the $p/d$ orbitals provides us the \textit{ab initio} tight-binding
Hamiltonian for computing the BC dipole~\cite{Fukui2005,BC_dipole}.

In Fig.~\ref{fig_1} we shows that the ML $\rm{WTe_2}$ with spin-orbit coupling~(SOC) exhibits a finite BC dipole. The SOC leads to an inverted, indirect quantum spin Hall gap.  In the DFT calculation, we used the Heyd-Scuseria-Ernzerhof~(HSE) method~\cite{Heyd_2003} with the hybrid parameter set at HSE$= 0.4$, which gives a global band gap of $50$ meV.    In the vicinity of the gap
minimum, the BC for the lowest conduction band exhibits two hotspots of opposite sign, as shown in Fig.~\ref{fig_1}(c). Such  bipolar configuration of BC is due to the $M_x$ symmetry.  It is worth noting that the distribution of BC near each hotspot is not uniform, leading to a nonzero BC dipole. 
To better understand the physics, we rewrite Eq.~\eqref{dipole} as $
D_x =  \sum_n \int d^2 \vec{k} \delta(\epsilon^{(n)}_{\vec{k}}-E_F)[ \partial_{k_x} \epsilon^{(n)}_{\vec{k}}] \Omega_{z}^{(n)} (\vec{k}),$ where $E_F$ is the Fermi energy and $\partial_{k_x} \epsilon^{(n)}_{\vec{k}}$ is associated with the velocity along $x$-direction. Let us first consider the case of one Dirac cone~(one hotspot). If  the BC is constant and the velocity is equal but opposite around a perfect Dirac cone, the BC dipole will vanish. However, for a tilted Dirac cone which has anisotropic BC and velocity around $E_F$, a finite BC dipole is allowed. In Fig.~\ref{fig_1} (c),  the red~(blue) hotspot shows more intense positive~(negative) BC in  the left~(right) part of the Dirac cone, where the velocity is negative~(positive) along x-direction, leading to a negative BC dipole.   For the highest valance band, the slope is negative along $\Gamma$ to $X$, as shown in Fig.~\ref{fig_1} (a). Thus, the configuration of BC in Fig.~\ref{fig_1} (d) gives a positive BC dipole. 

The $H$-structure monolayer TMDCs do not have nonlinear currents due to their $C_{3v}$ symmetry, but applying  uniaxial strain can reduce this symmetry
and leaving only a single mirror operation, in which case the quantum non-linear Hall effect can be observed~\cite{Sodemann2015}.  The application
of strain transforms the vector $\vec{r}_0$, which denotes undistorted crystal coordinate, into the new
position
$\vec{r} = \vec{r}_0 +\vec{u},$ where $\vec{r}_0=(x, y) $ and $\vec{u} = (u_x(x,y), u_y(x,y))$ are the position and  displacement deformation vector field, respectively. Here we are considering only the acoustic part of the in-plane displacement vector. In general, the derivative of $\vec{u}$ can be decomposed as $\vec\nabla \vec{u} = \varepsilon + \omega$, where $\varepsilon$ and
$\omega$ are the strain and rotation tensors, respectively.

\begin{figure}[t]
\centering
\includegraphics[width=8cm]{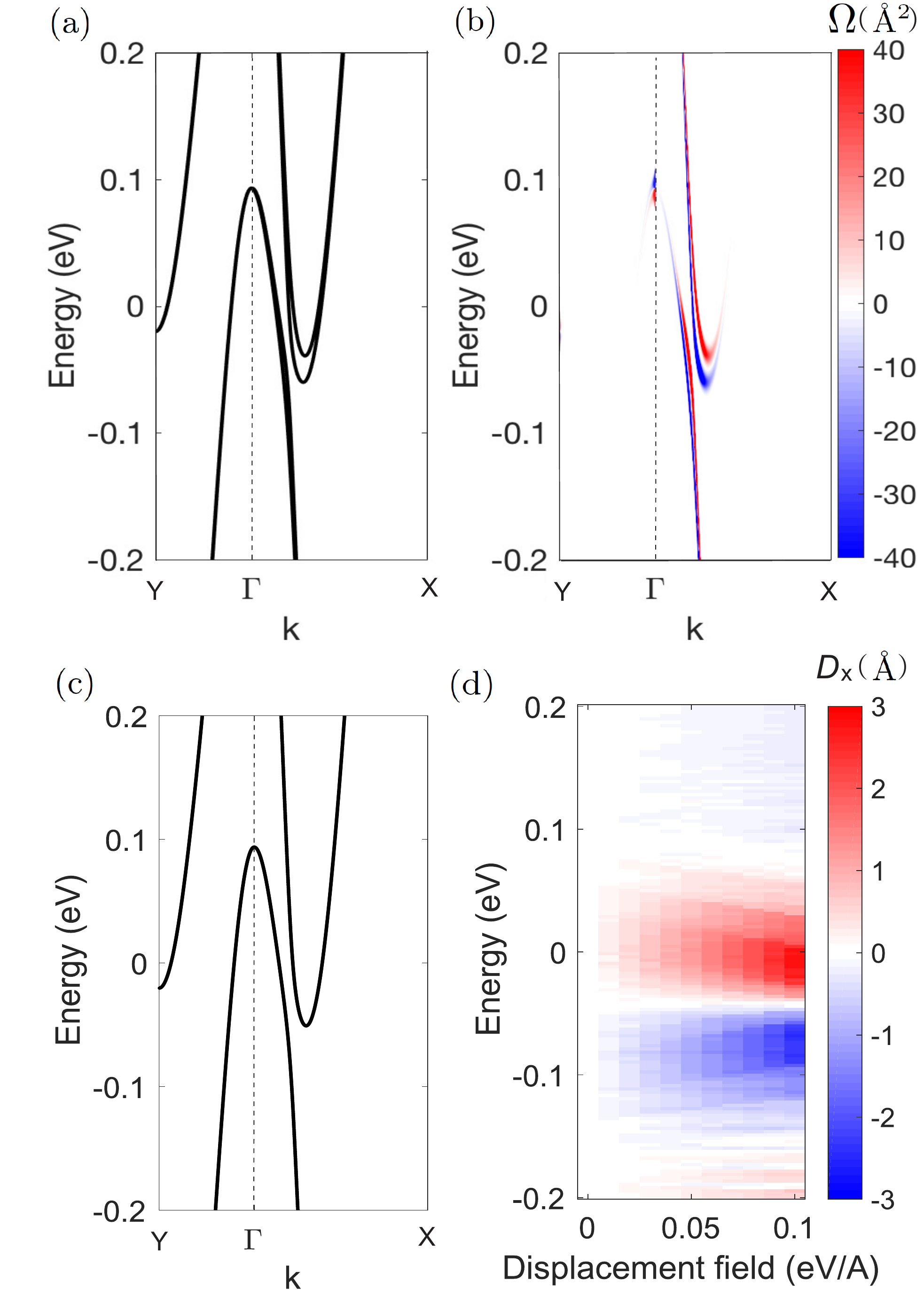}
\caption{(a) Band structure and (b) Berry curvature~$\Omega^{(n)}_z(\vec{k})$ along  Y$-\Gamma-$X  for the ML $1T'$-structure $\rm{MoTe_2}$ at $\mathcal{E}_{d,z}=0.06$ eV$/\rm{\AA}.$  (c) Band structure  at $\mathcal{E}_{d,z}=0.$ The BC is zero at all energies due to the presence of inversion symmetry. (d) BC dipole $D_x$  induced by the  electrical displacement field~(eV$/\rm{\AA}$) in the \emph{z}-direction which breaks the inversion symmetry. $D_y=0$ in all cases. 
}\label{fig_3}
\end{figure}

The simplest way to incorporate the effect of strain is to vary the interatomic
bond lengths, $|\delta_{\alpha\beta}|,$ between $\alpha$ and $\beta$ sites, which is known as the central force approximation. At the linear
order and under this approximation, the modified hopping terms in the presence of
strain can be approximated as
\begin{eqnarray}
t_{\alpha\beta}=t^0_{\alpha\beta}+\mu \vec{\delta}_{\alpha\beta}\cdot(\vec{\delta}_{\alpha\beta}\cdot \vec\nabla)\vec{u},
\mu=\frac{1}{|\delta_{\alpha\beta}|}[\frac{d t_{\alpha\beta}}{d |\delta_{\alpha\beta}|}].
\end{eqnarray}
The
central
force approximation  fails to capture the change in the hopping when the crystal is stretched along a
direction perpendicular to the bond.
The microscopic models based on the \textit{ab initio} derived Wannier functions describe up to linear order contributions in the strain  $(u_{xx} + u_{yy}), (u_{xx}- u_{yy})$ and $u_{xy}$ with respect to the
crystal symmetry and local crystal configuration, free
from any empirical fitting procedures.

Based on the models of strained TMDCs~\cite{Fang2017},
the BC dipole can be readily evaluated.
 Fig.~\ref{fig_2} shows that the $H$-structure $\rm{WSe_2}$ with strain exhibits BC dipole. The Brillouin zone of monolayer TMDC with Dirac
points is shifted away from the $K$ and $K'$ by  uniaxial  strain. When
the shear strain is applied along high-symmetry lines, one obtains finite $D_x$ but zero $D_y.$

The application of an out-of-plane  electrical displacement field can be used to systematically control the magnitude of
the nonlinear Hall current.
This can be understood from a third order
susceptibility tensor $\chi^{(3)}_{ijj z}$. In the presence of a static  electrical  displacement field $\mathcal{E}_{d,z}$ along the \emph{z}-direction the nonlinear dc current is written as $j^{(0)}_i = \chi^{(2)}_{ijj} |\mathcal{E}_j|^2+\chi^{(3)}_{ijj z} |\mathcal{E}_j|^2 \mathcal{E}_{d,z}.$ We obtain the effective second-order tensor containing the  electrical displacement field effect as $\tilde{\chi}^{(2)}_{ijj}\equiv \chi^{(2)}_{ijj}+\chi^{(3)}_{ijj z} \mathcal{E}_{d,z}.  $ When $\chi^{(2)}_{ijj}=0$ due to the intrinsic inversion symmetry, $\mathcal{E}_{d,z}$ makes it possible to
produce the second response with significant amplitude.

Finally we consider the basic symmetry properties of the
Hamiltonian for the TMDC family of
materials in the presence of  $\mathcal{E}_{d,z}$.
The $1T'$ phase has inversion symmetry from the combination of two-fold screw rotational symmetry $C_{2x}$ and  the mirror symmetry $M_x.$ 
 Therefore the monolayer $1T'$ structure, which has zero BC dipole, can acquire non-zero dipole if $\mathcal{E}_{d,z}$ is applied to break the $C_{2x}$ symmetry and thus inversion symmetry. This
inversion symmetry-breaking scheme can be simply modeled with only electrostatic
on-site potential within each unit cell. We show the  monolayer  $\rm{MoTe_2}$ with $1T'$ structure as a function of  electrical displacement $\mathcal{E}$-field in the \emph{z}-direction in Fig.~\ref{fig_3}. The enhancement of the BC dipole is evident. This illustrates how to control and modulate the BC dipole with an external  field.

In conclusion, 
we discussed the non-linear current induced by the BC dipole in 2D materials of the TMDC family. The existence of only a single mirror symmetry line is needed to provide a finite BC dipole. Breaking of the crystal symmetry can
be controlled by mechanical and/or electrical means.    
Certainly, it would be desirable to explore the proposed effect could be observed in other 2D materials subject to the same symmetry constraints.
Such a tunable BC dipole not only can lead to
the quantum nonlinear Hall effect but could also be relevant for the understanding of other quantum geometrical phenomena.

\begin{acknowledgements}
We thank Inti Sodemann, Bertrand I. Halperin, Philip Kim and Jeroen van den Brink for useful discussions. J.-S.Y. thanks Ulrike Nitzsche for technical assistance. This work was supported by the STC Center for Integrated Quantum Materials, NSF Grant No. DMR-1231319 and by ARO MURI Award W911NF-14-0247. The computations in this paper were run on the Odyssey cluster supported by the FAS Division of Science, Research Computing Group at Harvard University.  

Note added: Upon the completion of this manuscript, there appeared an independent work by Yang Zhang \emph{et al.}~\cite{Zhang2018_1} which also shows the nonlinear electric response in $\rm{MoTe_2}$  and $\rm{WTe_2}$ monolayers.
\end{acknowledgements}

\bibliographystyle{apsrev4-1}
\bibliography{berry_dipole,spin32,hetero}
\end{document}